\documentclass[prb,twocolumn,superscriptaddress]{revtex4-1}
\usepackage{amsthm}
\usepackage{amsmath}
\usepackage{graphicx}
\usepackage{bm}
\usepackage{subfigure}
\usepackage{tikz}
\usetikzlibrary{calc,3d}
\usetikzlibrary{arrows}
\usepackage[unicode=true,pdfusetitle,
 bookmarks=true,bookmarksnumbered=false,bookmarksopen=false,
 breaklinks=false,pdfborder={0 0 1},backref=false,colorlinks=false]
 {hyperref}
\hypersetup{
 linkcolor=blue}

\makeatletter

\usepackage{color}
\usepackage{transparent}

\makeatother

\begin{document}

\title{Spin-transfer torque induced spin waves in antiferromagnetic insulators}

\author{Matthew W. Daniels}

\thanks{These authors contributed equally.}

\affiliation{Department of Physics, Carnegie Mellon University, Pittsburgh, Pennsylvania
15213, USA}

\author{Wei Guo}

\thanks{These authors contributed equally.}

\affiliation{Department of Physics and State Key Laboratory of Surface Physics,
Fudan University, Shanghai 200433, China}

\author{G. Malcolm Stocks}

\affiliation{Materials Science and Technology Division, Oak Ridge National Laboratory,
Oak Ridge, Tennessee, 37831, USA}

\author{Di Xiao}

\email{dixiao@cmu.edu}

\affiliation{Department of Physics, Carnegie Mellon University, Pittsburgh, Pennsylvania
15213, USA}

\author{Jiang Xiao}

\email{xiaojiang@fudan.edu.cn}

\affiliation{Department of Physics and State Key Laboratory of Surface Physics,
Fudan University, Shanghai 200433, China}

\begin{abstract}
We explore the possibility of exciting spin waves in insulating antiferromagnetic films by injecting spin current at the surface. We analyze both magnetically compensated and uncompensated interfaces. We find that the spin current induced spin-transfer torque can excite spin waves in insulating antiferromagnetic materials and that the chirality of the excited spin wave is determined by the polarization of the injected spin current. Furthermore, the presence of magnetic surface anisotropy can greatly increase the accessibility of these excitations. 
\end{abstract}
\maketitle

\section{Introduction}

The field of spintronics seeks to investigate and organized phenomena 
concerning spin angular momentum. Of much recent interest in this field are 
spin-transfer torque (STT)~\cite{slonczewski_current-driven_1996, 
berger_emission_1996}, spin pumping,~\cite{tserkovnyak_enhanced_2002}, current
-induced magnetization dynamics (spin waves), \cite{kiselev_microwave_2003} 
the (inverse) spin Hall effect,~\cite{dyakonov_possibility_1971, 
hirsch_spin_1999} and the more recent spin 
caloritronics.~\cite{uchida_observation_2008} As a scientific enterprise, 
this rich intersection of spintronic physics hosts a vast and non-trivial 
dynamical landscape with many yet-unexplored avenues of research. Meanwhile, 
the degree to which spintronics can be applied to problems in computational 
information architecture is already quite promising, and it is likely that 
the full extent of these technologies is presently unrealized. Spintronics as 
a technological program ultimately means to provide a high-information-density
, energy-efficient computational architecture. STT and spin pumping provide a 
means to exchange spin into and out of a system, essentially constituting a I/
O layer for applications. Therefore, their study is central in connecting any 
spin-based computing scheme to a realistic electronic device.

Over the past two decades, scientists expended considerable effort was in 
learning to manipulate and detect ferromagnetic order via 
STT~\cite{slonczewski_current-driven_1996,berger_emission_1996} and spin 
pumping. ~\cite{tserkovnyak_enhanced_2002} This capacity to manipulate the 
magnetic order in ferromagnets---and, therefore, to initiate spin waves in 
the magnetic tecture---is the foundation of \emph{magnonics}. In the 
magnonics program, spin waves provide a complete replacement for itinerant 
electrons; information is no longer carried in conjunction with a flowing 
charge, but as a quasiparticle excitation of the background texture. The 
result is that no Joule heating is produced, making magnonics an attractive 
form of low-power computing. Recently, Kajiwara \emph{et al.}~have 
demonstrated the transmission of magnonics information, written and read via 
STT and spin pumping, in yttrium iron garnet 
(YIG).~\cite{kajiwara_transmission_2010} In their experiment, however, the 
critical current required to excite a spin wave was lower than expected. This 
descrepency was later resolved by Xiao \emph{et al.}, who theorized that the 
experimental apparatus had excited surface modes, rather than the expected 
bulk waves, and furthermore showed that these surface spin waves are 
associated with the considerably lower excitation threshold found in the 
experimental data.~\cite{xiao_spin-wave_2012}

Here, we are interested not in YIG, but in antiferromagnetic (AFM) magnonics. Because AFMs 
lack a net magnetization, their magnetic order is difficult to detect and 
control with magnetic fields. Applications of AFMs are consequently scarce, 
limited mostly to their use for exchange bias pinning of ferromagnetcs. 
Though spintronics research has historically focused on ferromagnetism, many 
theoretical works have considered how spin currents flowing through AFMs 
could interact with that magnetic order.~\cite{nunez_theory_2006,haney_current-induced_2008,xu_spin-transfer_2008,gomonay_spin_2010,linder_controllable_2011,swaving_current-induced_2011,hals_phenomenology_2011,cheng_electron_2012,tveten_staggered_2013,cheng_dynamics_2014,saidaoui_spin_2014} These studies of spintronics in AFM metals, however, address the electron as the spin carrier. The ideal system for magnonics would be in an insulator, where magnons alone are the dominant spin carrier. But until recently, there existed a colloquial understanding that AFM insulators could not support magnons with a nonzero spin, for any spin carried on one sublattice would be canceled by the other.

Recently, Cheng \emph{et al.}~shattered this illusion by showing that, in 
easy-axis AFMs, spin waves necessarily carry a spin angular momentum by 
adopting either a left- or right-handed chiral mode.\cite{cheng_spin_2014} 
Furthermore, they derived the magnetization dynamics due to STT in AFM 
insulators, and showed that both STT and spin pumping exist in AFMs and 
operate in a similar way to the ferromagnetic case. Despite these theoretical 
successes, some barriers still exist to realizing AFM magnonics in 
experiment. The most obvious dilemma is that, due to the strong exchange 
coupling in AFMs, the resonance frequency of bulk spin wave modes can be 
significantly higher than in ferromagnets---typically, it lies in the THz 
regime. Generating a THz signal is presently impossible by electronic means, 
as it would typically require a current which would melt the device before 
producing any meaningful effect.

In this article, we address the possibility of lowering the effective 
excitation threshold in AFM magnonics by considering the surface spin wave 
modes of AFMs. Our prediction relies on the fact that surface atoms in 
certain antiferromagnets can have an effective exchange energy significantly 
lower than the bulk value. One would then expect that the resonant frequency 
of spin waves localized to these exchange-reduced atomics will have lower 
excitation thresholds and be easier to excite. We compute the surface spin 
wave spectra of antiferromagnetic insulators with both magnetically 
compensated and uncompensated surfaces, and show that these surface modes 
are, as expected, lower in energy. We then include a contribution from STT 
due to a polarized spin current injected at the surface. We find that this 
STT is sufficient to excite the surface spin waves in low-surface-exchange 
systems, which demonstrates a step toward making AFM magnonics more 
realizable in experiment. We also show that the sign of the STT determines 
the handedness of the chiral AFM spin wave as predicted by Cheng \emph{et 
al}.\cite{cheng_spin_2014}

This paper is organized as follows: in Sec.~\ref{sec:Macrospin-Model}, we 
present a pedagogical model demonstrating not only that AFM spin waves can be 
excited by spin current, but also that the chirality of the spin wave depends 
on the spin current polarization. The distinction between source
chiralities of spin waves due to their oppositely carried angular momentum
could markedly improve the fidelity of devices utilizing 
the magnetization domain for information processing. In 
Sec.~\ref{sec:Lattice-Calculation}, we make the system more realistic by 
extending the dynamical equations from two sites to a full cubic lattice. 
Here we present new results on AFM spin wave spectra for a semi-infinite system 
with an interface. Based on previous work, \cite{xiao_spin-wave_2012} we 
expect that surface effects, by their role in modifying the magnons' 
activation threshold, will play an important part in the experimental 
realization of spin wave modes. In particular, we explore variations of the 
exchange coupling on the surfaces with both compensated and uncompensated net 
magnetizations. In Sec.~\ref{sec:Discussion}, we offer concluding remarks and 
an application for experimental methods.

\section{Macrospin Model\label{sec:Macrospin-Model}}

\begin{figure}
\centering
\includegraphics[width=\columnwidth]{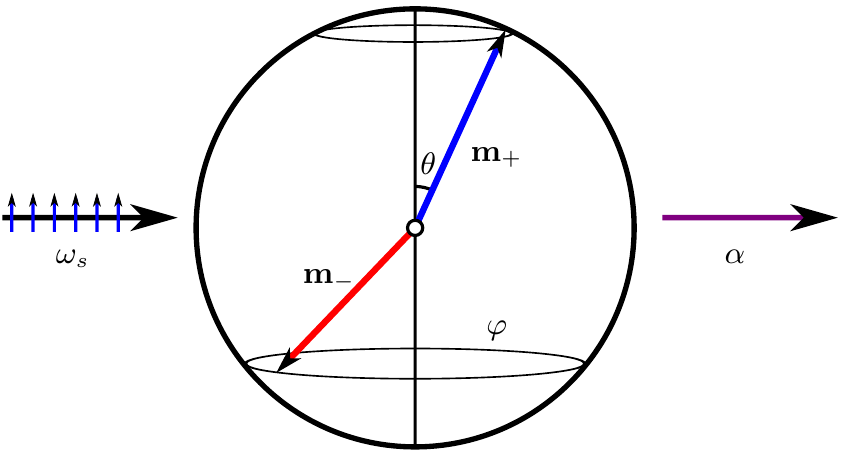}

\caption{(Color online) Cartoon setup for the macrospin model: two oppositely-oriented magnetic moments experience an exchange coupling $\omega_{J}$ via an on-site effective field. Spin-torque is injected via a polarized spin current $\omega_{s}$, and energy is dissipated via the Gilbert damping $\alpha$.  Not pictured are the external field and uniaxial anisotropy, both along $\hat{z}$.\label{fig:Setup-for-macrospin}} 
\end{figure}

To provide a conceptual account of the mechanism underlying STT-generated spin
waves in AF materials, we first present a minimal model which includes the 
important physical terms ithout the complications of a spatially extended 
lattice. This so-called ``macrospin model'' has been considered by many in the 
past, and we include it here not as new work but as a pedagogical
tool to illustrate the mechanism by which STT determines AFM spin wave
handedness. We will then extend this idea to our core result on a semi-infinite
lattice in Sec.~\ref{sec:Lattice-Calculation}.

In the macrospin mode, the magnetization on the two sublattices are modeled as
two macrospins, \cite{xiao_macrospin_2005} $\mathbf m_+$ and $\mathbf m_-$, 
which are coupled by a constant Heisenberg-type exchange interaction 
$\omega_{J}$. They are additionally subject to Gilbert damping $\alpha$ and 
spin transfer torque $\omega_{s}$; the latter is due to an injected spin 
current polarized along the $\hat{z}$ direction. Both macrospins experience
the a uniaxial anisotropy $\omega_{A}$ in the $\hat{z}$ direction. 
\cite{keffer_theory_1952} We also allow for an external magnetic field
\textbf{$H_{0}\hat{z}$} along this axis. The setup is depicted schematically
in Fig.~\ref{fig:Setup-for-macrospin}, and yields an equation of motion
\begin{align}
\dot{\mathbf{m}}_{\pm} 
&=-\mathbf{m}_{\pm}\times\mathbf{H}_{\text{eff}}^{\pm}
+\alpha\mathbf{m_{\pm}}\times\dot{\mathbf{m}}_{\pm}
+\omega_{s}\mathbf{m}_{\pm}\times
\left(\hat{z}\times\mathbf{m}_{\pm}\right) \nonumber \\
& =-\mathbf{m}_{\pm}\times\left[-\omega_{J}\mathbf{m}_{\mp}
+\left(\omega_{A}m_{\pm,z}+\omega_H\right)\hat{z}\right]\nonumber \\
&\phantom{=}+\alpha\mathbf{m_{\pm}}\times\dot{\mathbf{m}}_{\pm}+\omega_{s}\mathbf{m}_{\pm}\times\left(\hat{z}\times\mathbf{m}_{\pm}\right).\label{eq:LLG-macrospin}
\end{align}
where $\omega_{H}=\gamma H_{0}$. The effective field term $\mathbf{H}_{\text{eff}}^{\pm}$ is the negative derivative $-\nabla_{\mathbf{m}_{\pm}}H$ of the Hamiltonian
\begin{align}
H=\omega_{J}\mathbf{m}_{+}\cdot\mathbf{m}_{-}-\omega_{H}\left(m_{+}^{z}+m_{-}^{z}\right)-\frac{\omega_{A}}{2}\left({m_{+}^{z}}^{2}+{m_{-}^{z}}^{2}\right)\label{eq:macrospin-hamiltonian}
\end{align}
where the damping term is added phenomenologically, \cite{roepke_derivation_1971} and the
STT term---where $\omega_{s}$ is linear in the applied spin voltage---is due originally to
Ref.~\onlinecite{cheng_spin_2014}; we partially rederive it for the reader in
in Appendix \ref{sec:Appendix-LLG}.

In the small angle approximation, we demand that the deviation $\theta$ of $m_{z}\hat{z}$ from $\mathbf{m}$ be small, so that $m_{z}=\cos\theta=1+O\left(\theta^{2}\right)$ and $m_{x,y}\propto\theta+O\left(\theta^{3}\right)$.  Now the $\hat{z}$-component of equation Eq.~\eqref{eq:LLG-macrospin} vanishes to order $O\left(\theta^{2}\right)$, and the problem is reduced to two effective dimensions in the $xy$\nobreakdash-plane. 
We can exchange these two real dimensions for a single complex one by defining the transverse magnetization $u\equiv m_{x}+im_{y}$ and rewriting Eq.~\eqref{eq:LLG-macrospin} in terms of this new variable.  We then employ a spin wave \emph{ansatz} $u_{\pm}=\mu_{\pm}e^{-i\omega t}$ which allows us to solve for the modes that satisfy equation Eq. (\ref{eq:LLG-macrospin}).  In the small-angle approximation, these eigenfrequencies of precession are 
\begin{equation}
\omega_{\pm}=\pm\omega_{0}-i\alpha\left(\omega_{J}+\omega_{A}\right)\left(1\mp\frac{\omega_{s}}{\alpha\omega_{0}}\right).
\label{eq:macrospin-analytic-eigenfreq}
\end{equation}
The resonant frequency in the absence of damping and STT is $\omega_{0}=\sqrt{\omega_{A}\left(\omega_{A}+2\omega_{J}\right)}$. In AFM, two degenerate modes with opposite chirality appear in Eq.~\eqref{eq:macrospin-analytic-eigenfreq}. This equation highlights the essential competition between STT and precessional damping: when the applied spin current is sufficiently strong, the second term in Eq.~\eqref{eq:macrospin-analytic-eigenfreq} becomes positive and selectively excites one of the $\omega_\pm$ modes depending on the sign of $\omega_s$. Therefore, spin waves with different chirality can be selectively excited according to the spin current polarization. This behavior is different from STT-induced FM dynamics, for which only one polarization of spin current can excite FM spin waves while the other polarization enhances damping instead.   
\begin{figure}
\includegraphics[width=\columnwidth]{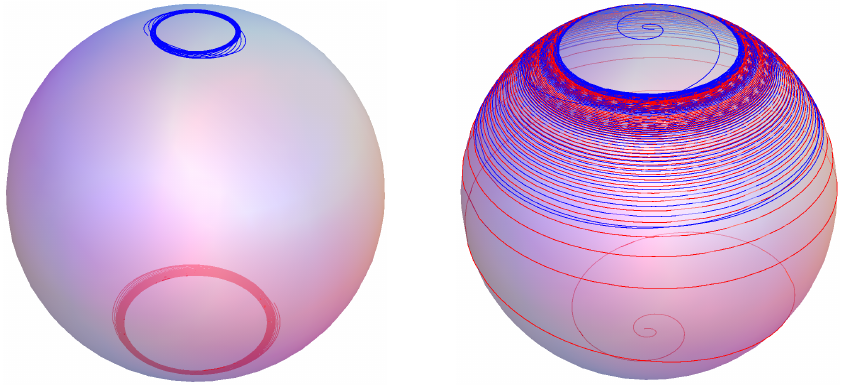}
\caption{(Color online) Left: With moderate $\omega_s$ which overcomes the damping effect, a stable oscillation AFM mode exists. This figure, which tracks the paths taken by spins $\mathbf{m}_{\pm}$ from Fig.~\ref{fig:Setup-for-macrospin}, is plotted for $t>100$ after the system has neared its steady state oscillation. Right: with stronger $\omega_{s}$, the system undergoes a spin flop, in which both $\mathbf{m}_\pm$ tilt to the north hemisphere. For both figures, $\omega_{A}/\omega_J=0.6$.\label{fig:A-stable-oscillation}}
\end{figure}

We can re-express $\mathbf{m}_{\pm}$ in spherical angular coordinates $\left(\theta^{\pm},\varphi^{\pm}\right)$ and derive a set of exact, coupled ODEs for this system directly from the coupled LLG equations. $\theta^\pm$ is taken to be the polar angle between $\mathbf{m}_{\pm}$ and $\hat{z}$, and $\phi^\pm$ is the corresponding azimuthal angle. We find
\begin{subequations}
\begin{align}
\dot{\theta}^{+} & =\omega_{J}\sin\Delta\varphi\sin\theta^{-}-\left(\alpha\dot{\varphi}^{+}+\omega_{s}\right)\sin\theta^{+}\label{eq:thetadot-p}\\
\dot{\varphi}^{+} & =\omega_{H}+\omega_{A}+\omega_{J}\sin\theta^{-}\cot\theta^{+}\cos\Delta\varphi\nonumber \\
 & \phantom{=}-\omega_{J}\cos\theta^{-}+\alpha\dot{\theta}^{+}\csc\theta^{+}\label{eq:phidot-p}\\
\dot{\theta}^{-} & =-\omega_{J}\sin\Delta\varphi\sin\theta^{+}-\left(\alpha\dot{\varphi}^{-}+\omega_{s}\right)\sin\theta^{-}\label{eq:thetadot-m}\\
\dot{\varphi}^{-} & =\omega_{H}-\omega_{A}+\omega_{J}\sin\theta^{+}\cot\theta^{-}\cos\Delta\varphi\nonumber \\
 & \phantom{=}-\omega_{J}\cos\theta^{+}+\alpha\dot{\theta}^{-}\csc\theta^{-}\label{eq:phidot-m}
\end{align}
\end{subequations}
where $\Delta\varphi=\varphi^{+}-\varphi^{-}$. This result is analytically exact. Some numerical calculations for these ODEs are depicted in Fig.~\ref{fig:A-stable-oscillation}.  Since the exchange energy is locally minimized where $\varphi^{+}-\varphi^{-}=\Delta\varphi=\pi$, we expect $\dot{\varphi}^{+}=\dot{\varphi}^{-}$. In the small angle approximation and neglecting $\dot{\theta}_{\pm}$ terms, this condition is satisfied when
\begin{equation}
\frac{\vartheta^{+}}{\vartheta^{-}}=-\left(\frac{\omega_{J}+\omega_{A}}{\omega_{J}\cos\Delta\varphi}\right)\pm\sqrt{\left(\frac{\omega_{J}+\omega_{A}}{\omega_{J}\cos\Delta\varphi}\right)^{2}-1}.\label{eq:theta-ratio-macromode}
\end{equation}
where $\vartheta^{\pm}$ are the angles that $\mathbf{m}_{\pm}$ make
with the $\pm\hat{z}$ axes. Choosing $\Delta\varphi=\pi$, as energetically
expected, recovers the results from Ref. \onlinecite{keffer_theory_1952}.
Within the $\dot{\theta}\approx0$ approximation, there is no real 
solution for $\vartheta^{+}=\vartheta^{-}$ in the presence of
easy-axis anisotropy, and one spin will always dominate the dynamics.
Because the spins stay antiparallel, the two chiral modes correspond to
a right-handed or left-handed rotation of the (+)-sublattice, and always
carry a net angular momentum.\cite{cheng_spin_2014}
\section{Lattice Calculation\label{sec:Lattice-Calculation}}

\begin{figure*}[t]
\includegraphics[width=2\columnwidth]{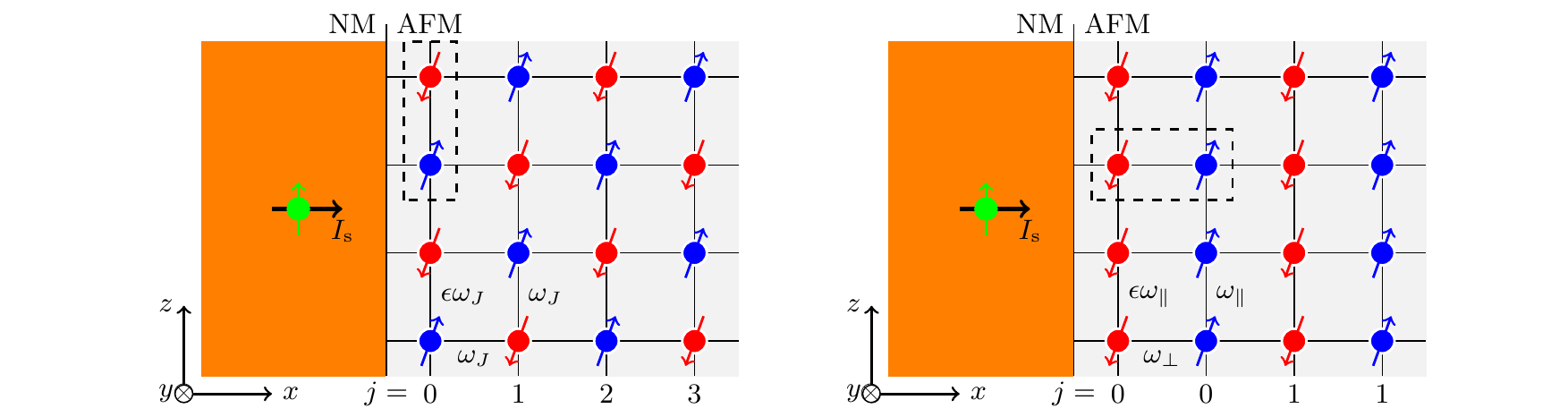}
\caption{(Color online) 2D slices of the spin configurations for $g$-type (left) and $a$-type (right) AFM, interfacing with NM. For $g$-type, the neighboring spins in the bulk have exchange coupling $\omega_J$, but have coupling $\epsilon\omega_J$ on the surface. For $a$-type, the intralayer exchange coupling is $\omega_{\|}$ in the bulk and $\epsilon\omega_{\|}$ on the surface, while the interlayer exchange coupling is $\omega_\perp$. In both cases, the far left column of spins is the $x=j=0$ atomic surface layer which sits against a nonmagnetic interface.  Unit cells are outlined in dashed box. Spin current $I_{\rm s}$ is injected from NM and exerting a torque on the surface spins.  
\label{fig:spin-texture-diagrams}}
\end{figure*}

%

To consider a more realistic system than that of Sec.~\ref{sec:Macrospin-Model},
we now extend the Heisenberg-type Hamiltonian \eqref{eq:macrospin-hamiltonian} 
to a simple cubic lattice as in Ref.~\onlinecite{wolfram_surface_1969}:
\begin{equation}
H=\sum_{\left\langle i,j\right\rangle }\omega_{ij}\mathbf{m}_{i}\cdot\mathbf{m}_{j}-\sum_{j}\left(\omega_{H}+\frac{\omega_{A}}{2}m_{j,z}\right)m_{j,z},\label{eq:heisenberg-hamiltonian}
\end{equation}
where the subscripts $i$ and $j$ are lattice sites and the first sum is 
taken over nearest neighbors.

We will consider both \emph{g}- and \emph{a}-type antiferromagnets. These 
configurations are depicted in Fig.~\ref{fig:spin-texture-diagrams}, where
the AFM terminates at $x=0$ with compensated (left) and uncompensated (right)
surfaces. We take the lattice constant as $a=1$ so that the wavevector is 
dimensionless.  The thermodynamic derivation from Sec.~\ref{sec:Macrospin-Model}
is repeated for the Hamiltonian in Eq.~(\ref{eq:heisenberg-hamiltonian}) to
derive an effective on-site magnetic field. Knowledge of this field determines
the LLG equation for $\dot{\mathbf{m}}_{j}$, namely:
\begin{equation}
\dot{\mathbf{m}}_{j}=-\mathbf{m}_{j}\times\left[\sum_{\left\langle i,j\right\rangle }\omega_{ij}\mathbf{m}_{i}-(\omega_{H}-\omega_{A}m_{j,z})\hat{z}\right]+\alpha\mathbf{m}_{j}\times\mathbf{\dot{m}}_{j}.
\label{eq:LLG-basic}
\end{equation}
The exchange coefficients $\omega_{ij}$ will be uniformly constant $\omega_{ij}=\omega_{J}$ for the \emph{g}-type system where all nearest neighbors are the same, though for the \emph{a}-type system we will need to distinguish $\omega_{ij}=\omega_{\perp}<0$ and $\omega_{ij}=\omega_{\parallel}>0$ for the coupling between inter- and intra-plane (respectively AFM-like and FM-like) neighbors.

By assuming a small precession of $\mathbf{m}_{j}$ about its easy-axis, the
$\hat{z}$-component of the LLG Eqs.~\eqref{eq:LLG-basic} can be neglected
to first order. We then rewrite the equation of motion in terms of the 
transverse magnetization $u^{\pm}\equiv m_{\pm,x}+im_{\pm,y}$ as in 
Sec.~\ref{sec:Macrospin-Model}. Translational symmetry in time and 
the $yz$\nobreakdash-plane validates the plane wave \emph{ansatz}
\begin{equation}
u_{(j,\mathbf{s})}^\pm=\mu^\pm_{j,\mathbf{q}}e^{i\left(\mathbf{q}\cdot\mathbf{s}-\omega t\right)}\label{eq:plane-wave-ansatz}
\end{equation}
where $j$ is the layer index in the $\hat{x}$-direction and $\mathbf{q}$ is the wave vector in $yz$-plane. We substitute this equality into the transverse magnetization equation. From now on we will use $\mathbf{k}$ to refer to a 3D wavevector and $\mathbf{q}$ will be $\mathbf{k}$'s restriction in the $yz$-plane. 

With these modifications, the LLG Eq.~\eqref{eq:LLG-basic} is rewritten as a recurrence relation among different layers
\begin{equation}
S\psi_{j}+N_{+}\psi_{j+1}+N_{-}\psi_{j-1}=0\label{eq:bulk-recurrence}
\end{equation}
with $\psi_{j}=\left(\mu_{j}^{+}\quad\,\mu_{j}^{-}\right)^{T}$.  The square matrices $S$, $N_{+}$, and $N_{-}$ can be computed directly from considering the coefficients in Eq. (\ref{eq:LLG-basic}).  For $g$-type,
\begin{subequations}
\begin{align}
S^{(g)} & =\left(\begin{matrix}\omega-\omega_{H}-\Omega & -\omega^{(g)}_{q}\\
\omega^{(g)}_{q} & \omega-\omega_{H}+\Omega
\end{matrix}\right),\label{eq:S-matrix-G}\\
N^{(g)}_{+} & =N^{(g)}_{-}=\left(\begin{matrix}0 & -\omega_{J}\\
\omega_{J} & 0
\end{matrix}\right)\label{eq:N-matrix-G}
\end{align}
\end{subequations}
with $\Omega=6\omega_{J}+\omega_{A}-i\alpha\omega$ and $\omega^{(g)}_{q}=2\omega_{J}\left(\cos q_{y}+\cos q_{z}\right)$.  For \emph{a}-type, 
\begin{subequations}
\begin{align}
S^{(a)} & =\left(\begin{matrix}\omega-\omega_{H}-\Omega_{q} & -\omega_{\perp}\\
\omega_{\perp} & \omega-\omega_{H}+\Omega_{q}
\end{matrix}\right),\label{eq:S-matrix-A}\\
N^{(a)}_{+} & =\left(\begin{matrix}0 & -\omega_{\perp}\\
0 & 0
\end{matrix}\right),\ N^{(a)}_{-}=\left(\begin{matrix}0 & 0\\
\omega_{\perp} & 0
\end{matrix}\right)\label{eq:N-matrix-A}
\end{align}
\end{subequations}
with $\Omega_{q}=2\omega_{\perp}+\omega_{A}+\omega_{q}^{(a)}-i\alpha\omega$ and $\omega^{(a)}_{q}=2\omega_{\parallel}\left(2-\cos q_{y}-\cos q_{z}\right)$.

\subsection{Bulk calculation\label{sub:Bulk-calculation}}

For comparison with our results for a semi-infinite lattice in
Sec.~\ref{sub:Framework-for-solving}, we pause to reproduce the bulk spin wave
spectrum this formalism. The reader may refer to
Ref.~\onlinecite{wolfram_surface_1969}, or to any condensed matter theory
textbook, for a more complete discussion of the g-type spectral calculation.
The a-type calculation is similar except that the primitive lattice vectors
differ.

In addition to the translational symmetries used in the previous section,
a bulk lattice possesses an additional 
translational symmetry in the $\hat{x}$ direction. Therefore we may take a plane wave 
solution for the $x$-coordinate: $\psi_j = \phi(q)e^{i(k_x j-\omega t)}$. We 
can then find the eigenfrequencies of AFM spin waves.  For \emph{g}-type, 
\begin{equation}
\omega_{(g)}=\omega_{H}\pm\sqrt{\Omega^{2}-\omega_{k}^{2}}\label{eq:g-type-bulk-dispersion}
\end{equation}
with $\omega_{k}=2\omega_{J}\left(\cos k_{x}+\cos k_{y}+\cos k_{z}\right)$.
For the \emph{a}-type lattice, we have an infinite stack of alternating
ferromagnetic sheets. We choose for $\hat{x}$ to be the direction normal to
any given sheet. Then the bulk dispersion is 
\begin{equation}
\omega_{(a)}=\omega_{H}\pm\sqrt{\Omega_{q}^{2}-4\omega_{\perp}^{2}\cos^{2}k_{x}}.\label{eq:a-type-bulk-dispersion}
\end{equation}
One can verify that in the limit $\omega_{\parallel}=0$ we recover decoupled 1D AF chains with the expected dispersion $\pm2\omega_{\perp}\left|\sin k_{x}\right|$ in the simple isotropic case. \cite{kittel_introduction_2005} Likewise, the $\omega_{\perp}=0$ limit recovers a decoupled 2D ferromagnetic system, with dispersion $\pm2\omega_{\parallel}\left|\cos k_{y}+\cos k_{z}-2\right|$ in the same simple case.

The spin wave eigenfunctions corresponding to Eqs.~(\ref{eq:g-type-bulk-dispersion},\ref{eq:a-type-bulk-dispersion}) are respectively
\begin{subequations}
\label{eq:phi}
\begin{align}
\varphi_{\pm}^{\left(g\right)} & =\left(\begin{matrix}\sqrt{\Omega+\left(\omega-\omega_{H}\right)}\\
\mp\sqrt{\Omega-\left(\omega-\omega_{H}\right)}
\end{matrix}\right),\label{eq:phiG}\\
\text{and }\varphi_{\pm}^{\left(a\right)} & =\left(\begin{matrix}e^{ik_x}\sqrt{\Omega\pm\left(\omega-\omega_H\right)}\\
-\sqrt{\Omega\mp\left(\omega-\omega_H\right)}
\end{matrix}\right).\label{eq:phiA}
\end{align}
\end{subequations}

The dispersion relations (\ref{eq:g-type-bulk-dispersion},\ref{eq:a-type-bulk-dispersion}) enforce a constraint linking the irreducible representations of time ($\omega$) and space ($k$) translational symmetries. For any particular $\omega$ and $\mathbf{q}$, there are only a finite number of $k_{x}$ values in its preimage under the eigenvalue Eqs.~(\ref{eq:g-type-bulk-dispersion},\ref{eq:a-type-bulk-dispersion}).  In the next section, we will need to consider linear combinations of bulk solutions to satisfy the boundary condition. The dispersion relations above will allow us to consider only a small subset of all conceivable wavenumbers $k_{x}$.

\subsection{The semi-infinite case\label{sub:Framework-for-solving}}

We now introduce an interface by terminating the insulator along its $(100)$ plane and replacing the space $x<0$ with a nonmagnetic contact from which spin current can be injected. We will modify the equations of motion to allow for special conditions on the atomic surface layer at $x=0$.

\begin{figure*}[t]
\centering
\includegraphics[width=1.9\columnwidth]{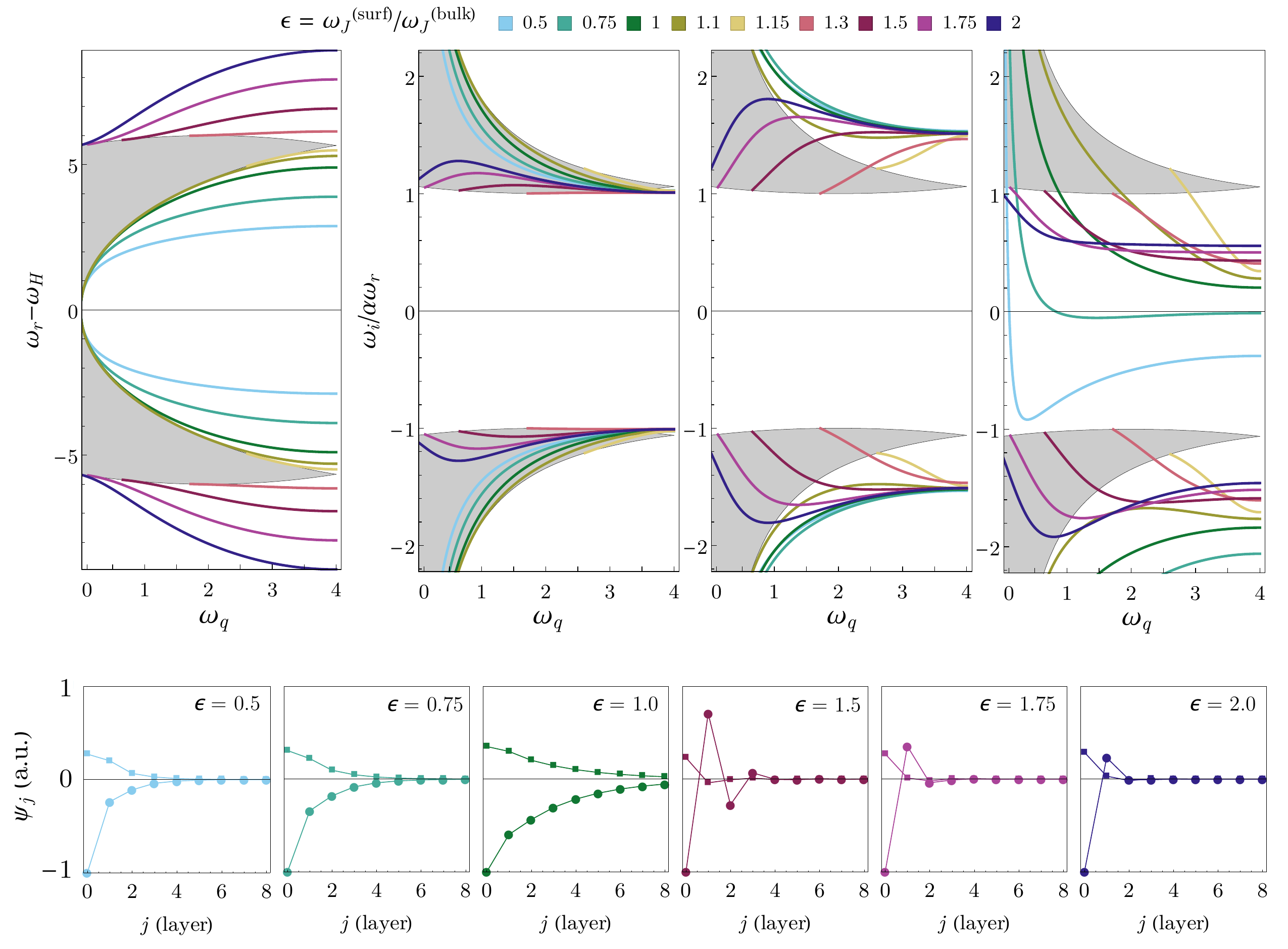}
\caption{(Color online) Frequency dispersion for the \emph{g}-type
semi-infinite system. From top left to top right: $\omega_r$ and then
$\omega_i/\alpha\omega_r$ for $\alpha=0.01$; $\alpha=0.01$ and $\beta=\alpha/2$;
$\alpha=0.01$ and $\omega_{s}=4\alpha$.  The gray regions indicate the bulk spectrum. The
horizontal axes measure $\omega_q \equiv 4-\omega_{q}^{(g)}$, so that $\omega_q=0$
corresponds to the $\Gamma$ point in the surface Brillouin zone. An array of spin wave profiles plotting 
the magnitudes of $\psi_j^\pm$ at $4-\omega_q^{(g)} = 1.5$ is shown in the bottom row.
\label{fig:Eigenfrequencies-dispersion-for-gtype-semi-infinite}}
\end{figure*}

\begin{figure*}[t]
\centering
\includegraphics[width=1.9\columnwidth]{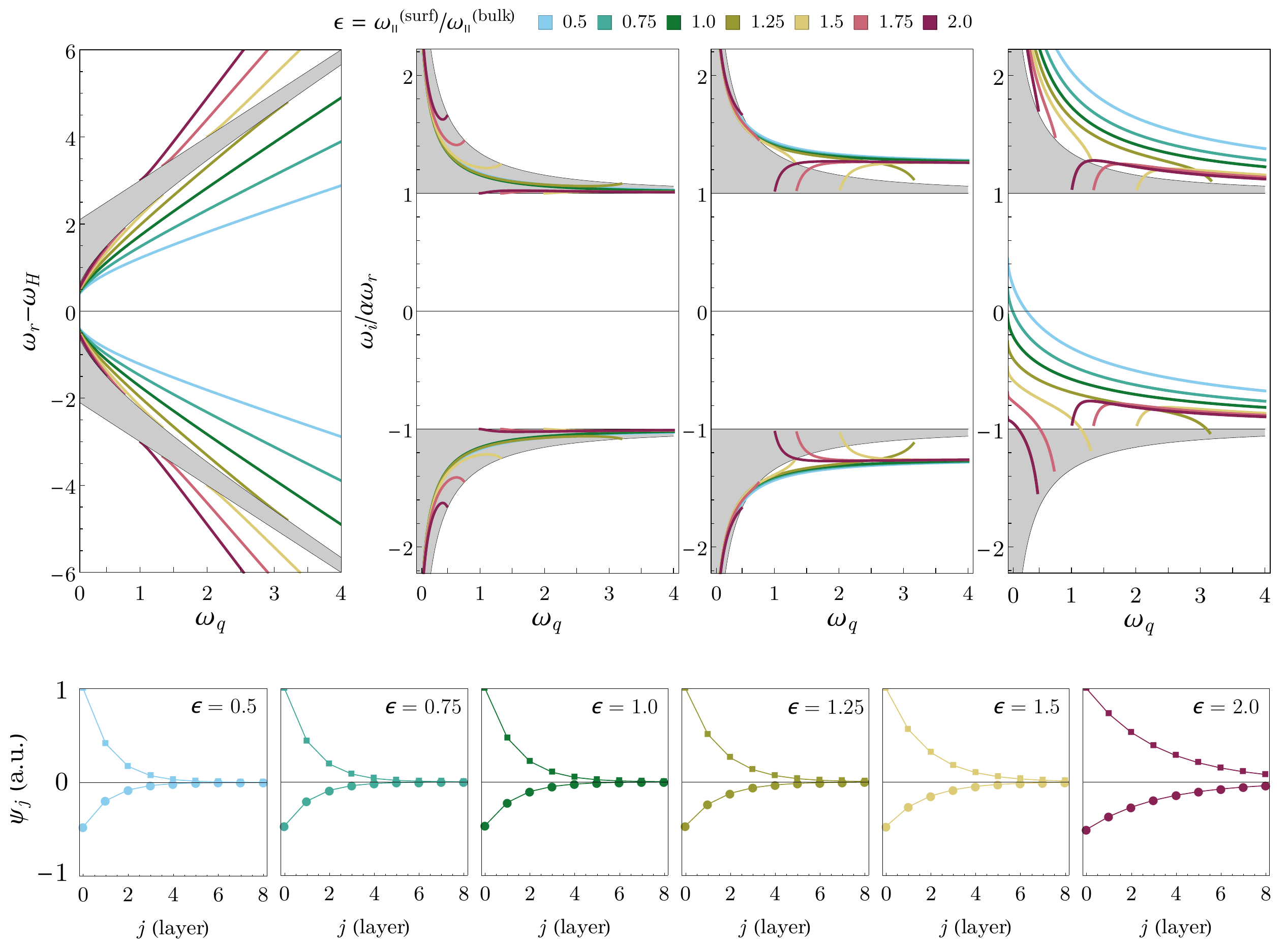}
\caption{(Color online) Frequency dispersion for the \emph{a}-type
semi-infinite system with $(\omega_\|, \omega_\perp) = (0.5,1)$. From left to
right: $\omega_r$ and then $\omega_i/\alpha\omega_r$ for $\alpha=0.01$;
$\alpha=0.01$ and $\beta=\alpha/4$; $\alpha=0.01$ and $\omega_{s}=\alpha$. The gray regions indicate the bulk
spectrum. The horizontal axes measure $\omega_q \equiv \omega_{q}^{(a)}$, so that $\omega_q=0$
corresponds to the $\Gamma$ point in the surface Brillouin zone. An array of spin
wave profiles plotting the magnitudes of $\psi_j^\pm$ at $\omega_q^{(a)} = 0.3$ is shown below.
\label{fig:Eigenfrequencies-dispersion-for-atype-semi-infinite}}
\end{figure*}

First, an enhanced damping term is inserted into the LLG equation by taking
$\alpha\mapsto\alpha+\beta\delta_{j\pm,0}$, where $\beta$ is the enhanced 
damping parameter for the surface spins. This enhanced damping represents 
spin loss due to the spin pumping effect from the AFM back into the NM
contact. The STT term 
$\omega_{s}\mathbf{m}_{j}\times\left(\hat{\mathbf{z}}\times\mathbf{m}_{j}\right)\delta_{j\pm,0}$
is likewise included on the atomic surface layer.  Finally, as a form 
of surface anisotropy, we allow a modulation of the intralayer exchange 
coupling represented by the ratio
$\epsilon\equiv\omega_{J}^{\text{surf}}/\omega_{J}^{\text{bulk}}$
(or $\omega_{\parallel}^{\text{surf}}/\omega_{\parallel}^{\text{bulk}}$).
It is known that this type of surface anisotropy can induce surface 
spin wave modes in AFM. \cite{wolfram_surface_1969}  The variation in the
exchange energy at the interface of magnetic materials has been studied 
by many groups. For instance, numerical studies on NiO(100) interfaces 
have shown that, depending on the assumptions of the model, surface 
exchange energy can vary by at least 20\% with some groups showing as 
much as a 50\% variation \cite{kodderitzsch_exchange_2002} from the bulk coupling.

We can write new equations of motion for this semi-infinite system as:
\begin{subequations}
\begin{align}
\left(S+B\right)\psi_{0}+N_{+}\psi_{1} & =0\quad\left(j=0\right)\label{eq:recursive-base-case}\\
S\psi_{j}+N_{+}\psi_{j+1}+N_{-}\psi_{j-1} & =0\quad\left(j>0\right)\label{eq:recursive-induction}
\end{align}
\end{subequations}
where, for \emph{g}- and \emph{a}-types, respectively:
\begin{subequations}
\begin{align}
&B^{(g)} = \left[\left(5-4\epsilon\right)\omega_{J}+i\left(\omega_{s}+\beta\omega\right)\right]\sigma_{z}+\omega_{q}\left(1-\epsilon\right)i\sigma_y\label{eq:boundary-G}, \\
&B^{(a)}=-\left[\omega_{\perp}+\omega_{q}\left(1-\epsilon\right)-i\left(\omega_{s}-\beta\omega\right)\right]
\frac{1-\sigma_{z}}{2},\label{eq:boundary-A}
\end{align}
\end{subequations}
and $\sigma_{x,y,z}$ are the Pauli matrices.

We now take the bulk eigenvectors $\varphi_{\pm}$ in Eq. (\ref{eq:phi}) for the $g$-type as a basis for general solutions to a semi-infinite lattice configuration. By using the bulk dispersion relations, $\varphi_{\pm}$ can be rewritten in terms of a distinguished eigenvalue $\omega$ and trigonometric functions of $k_{x}$ as in Eq. (\ref{eq:phi}). Recall from the conclusion of Section \ref{sub:Bulk-calculation} that for a particular value of $\omega=\omega(\mathbf{q})$ the irreducible representation $k_{x}$ is restricted to the finite set of values $k_{x}\in\omega_{(g,a)}^{-1}(\omega\left(\mathbf{q}\right))$.  We will call these at most four values by $k_{\mu}$. Since the cosine function is even, we see that two of the $k_{\mu}$ values are related by a sign change to the other two. As will become clear in Sections \ref{sub:Compensated-surface} and \ref{sub:Uncompensated-surface}, we demand that $\Im\left(k_{x}\right)$ be positive so that surface solutions decay into the bulk. Then two allowed values of $k_{\mu}$ remain, which we call $k^{+}$ and $k^{-}$.

We can now consider solutions of the form
\begin{equation}
\psi_{j}=\eta_{+}\varphi_{+}e^{i(k^{+}j-\omega t)}+\eta_{-}\varphi_{-}e^{i(k^{-}j-\omega t)}\label{eq:g-type-general-solution}
\end{equation}
where $\varphi_{\pm}$ are the bulk eigenvectors corresponding to $k^{\pm}$, which are the only allowed wavenumbers $k_{x}$ in the preimage of the bulk $\omega$.

\subsubsection{\emph{g}-type, with compensated surface}
\label{sub:Compensated-surface}

With Eq.~\eqref{eq:g-type-general-solution}, the boundary condition Eq. (\ref{eq:recursive-base-case}) for the compensated $g$-type system takes the form
\begin{equation}
\det\left[B\left(\varphi_{+}\;\varphi_{-}\right)+N\left(\varphi_{+}e^{ik^{+}}\;\varphi_{-}e^{ik^{-}}\right)\right]=0.\label{eq:g-type-det}
\end{equation}
The exponentials $e^{ik^{\pm}}$ can be determined from solving the eigenvalue equations Eq. (\ref{eq:g-type-bulk-dispersion},\ref{eq:a-type-bulk-dispersion}) for $\cos k_{x}$, employing the Pythagorean identity to expand Euler's formula, and demanding solutions $\Im\left(k_{x}\right)>0$ which decay into the bulk. Taken together with Eq. (\ref{eq:g-type-bulk-dispersion}), this equation can be solved analytically for $\omega$ when $\alpha=\beta=\omega_{s}=0$.  This unperturbed eigenfrequency is then used to calculate constant perturbations---namely the $i\alpha\omega$ and $\beta\omega$ terms---so that equation Eq. (\ref{eq:g-type-det}) can be evaluated to leading order in the presence of damping and STT with straightforward modifications to its coefficients.  The results in the complex eigenfrequencies $\omega = \omega_r + i\omega_i$ are plotted in Fig.~\ref{fig:Eigenfrequencies-dispersion-for-gtype-semi-infinite}, wherein the bulk modes are plotted as the shaded area and the surface modes are plotted in colored curves. To the leftmost panel of Fig.~\ref{fig:Eigenfrequencies-dispersion-for-gtype-semi-infinite} corresponds the real part of the eigenfrequency $\omega_r - \omega_H$ (in units of $\omega_J$), and the right three panels are the imaginary part $\omega_i/\alpha\omega_r$ for three different cases: purely intrinsic damping, with neither spin pumping nor STT; both damping and spin pumping (due to the enhanced damping, $\beta$), but no STT; and both damping and STT, but no spin pumping.  

The dispersion relations of $\omega_r$ for the surface modes of this system
are plotted in
Fig.~\ref{fig:Eigenfrequencies-dispersion-for-gtype-semi-infinite} over a
spectrum of surface exchange ratios $\epsilon$. These surface modes are the
same as those calculated in Ref. \onlinecite{wolfram_surface_1969}. The spin
wave profiles for the surface modes are presented in lower panels of the
figure, which shows a positive correlation between surface localization and
surface anisotropy. These figures also reveal that the surface modes in a
g-type AFM can be either acoustic or optical; a detailed discussion of the
acoustic/optical transition as a function of $\epsilon$ is given in
Ref.~\onlinecite{wolfram_surface_1969}.

Beyond the dispersion relations, we are also interested in the dissipative
behavior of various spin wave modes. Especially of interest are their
behavior under the influence a STT due to spin current injection from the NM
contact. The second panel of
Fig.~\ref{fig:Eigenfrequencies-dispersion-for-gtype-semi-infinite} shows
$\omega_i$ when there is only intrinsic damping included. In this case there
is neither spin pumping or STT, and we plot both the bulk modes (shaded
continuum) and the surface modes (colored curves) for different values of the
surface anisotropy $\epsilon$. With the additional NM contact at the surface,
the spin pumping into NM from AFM increases the dissipation for the spin wave
modes, as seen in the third panel in
Fig.~\ref{fig:Eigenfrequencies-dispersion-for-gtype-semi-infinite}. Far from
where the surface modes emerge from the bulk spectrum, the effective damping
enhancement (in the language of Ref.~\onlinecite{brataas_damping}) is
$\Delta\alpha\approx\beta$. This is expected since this regime corresponds
to high surface localization, wherein $\beta$ is effectively just added to
$\alpha$ in the local LLG equations.
introduction of a spin-transfer torque can dramatically decrease the damping
of some surface spin waves, especially in the low-$\epsilon$ regime where
surface anisotropy is strong. The low damping combined with low excitation
energy makes these low-$\epsilon$ modes particularly excitable due to strong
surface localization. Strong enough $\omega_{s}$ together with low $\epsilon$
(strong surface anisotropy) can cause sign changes in $\omega_i$ and lead to
AFM spin wave excitation, as in the last panel of
Fig.~\ref{fig:Eigenfrequencies-dispersion-for-gtype-semi-infinite}.
Furthermore, STT distinguishes the two spin wave chiralities by enhancing the
damping of one while reducing the other. Precisely which chirality is excited
depends on the spin current polarization, so that it is distinctly possible
to selectively excite a particular chiral mode.  



\subsubsection{\emph{a}-type, with uncompensated surface}
\label{sub:Uncompensated-surface}

For the uncompensated surface in an \emph{a}-type AFM insulator, there is
effectively only one $k_{x}$ which satisfies both the bulk eigenfrequency
equations and the reality condition $\Im\left(k_{x}\right)>0$ for any given
$\omega$. The reasoning follows: first, the orientation of the unit cell is
necessarily different in the \emph{a}-type system, so that the coupling to
the next unit cell along the $x$-direction requires a factor of $e^{2ik}$
rather than just $e^{ik}$ in the \emph{a}-type analog to equation
Eq.~\eqref{eq:g-type-general-solution}; second, solving equation Eq.
(\ref{eq:a-type-bulk-dispersion}) for $k_{x}$ gives a family of four
solutions---namely $k$, $-k$, $\pi+k$, and $\pi-k$---but as we mentioned in
Sec.~\ref{sub:Framework-for-solving}, only one of $k$ and $-k$ will have a
positive imaginary part, and they furthermore will each appear identical to
their $\pi$-shifted partners when expressed in the form $e^{2ik}$. 
This simplifies the form of the boundary condition
Eq.~\eqref{eq:recursive-base-case}, as well as the \emph{a}-type analog of
Eq.~\eqref{eq:g-type-det}. A similar procedure to that employed in the
previous section is used to solve the unperturbed and then perturbed versions
of this equation. 

The spin wave dispersion $\omega_r$ for an $a$-type AFM is different from
that for $g$-type AFM; this is evident in the left panel of
Fig.~\ref{fig:Eigenfrequencies-dispersion-for-atype-semi-infinite}. However,
the surface anisotropy still induces surface spin wave modes. Typical surface
mode profile are shown below the dispersion plots.  
In the absorption spectra (right three panels of Fig.~\ref{fig:Eigenfrequencies-dispersion-for-atype-semi-infinite}), the
spin pumping (third panel) enhances the dissipation for both chiralities
(again at $\Delta\alpha\approx\beta$)
while STT reduces the dissipation for one chirality and enhances the other.
These results coincide with the outcomes of Section
\ref{sub:Compensated-surface} for the \emph{g}-type configuration, again
distinguishing spin wave chiralities and demonstrating that a nonzero
$\omega_{s}$ in the \emph{a}-type system can cause a change in sign of the
absorption spectrum, and can consequently excite spin wave modes.

\section{Conclusions\label{sec:Discussion}}

In this article, we have calculated the spin wave spectrum of 
STT-induced AFM surface excitations. In particular,
we found that surface spin wave modes induced by surface anisotropy are
particularly easy to excite compared to bulk modes, implying a lowering
of the naive critical current needed to perform magnonic operations in
AFM insulators. 

As we noted in the Introduction, the efficiency of spin pumping processes in
antiferromagnets is known to be comparable to their ferromagnetic
cousins.\cite{cheng_spin_2014} However, because antiferromagnets have a much
stronger exchange coupling, an \emph{a priori} estimate of the threshold
current for exciting AFM surface spin waves is two to three orders of
magnitude higher than in ferromagnet insulators. Nevertheless, the critical
current for exciting a ferromagnetic magnon current was found in Ref.~\onlinecite{kajiwara_transmission_2010}
to be two to three orders of magnitude lower than the expectation accorded
to YIG's resonant frequency. If the same unforeseen reduction occurs in AFM,
then the critical current would be on the order $J_c\approx10^8 A/cm^2$, which is
within experimental feasibility. Our contribution is to take a first step
in investigating this potential reduction in the critical barrier. 
One may of course seek materials with appropriate exchange or anisotropy energies in accordance with
Eq.~\eqref{eq:macrospin-analytic-eigenfreq} in order to lower the barrier;
we find that seeking materials with low surface exchange coupling reduces
the threshold further.

Our work also takes a first step toward developing new experimental techniques
for investigating antiferromagnets. Because it is relatively straightforward
to generate a spin current and measure spin waves, STT-based methods could 
provide a new tool for probing and controlling AF materials. In particular, 
parameters such as damping, anisotropy, or surface exchange coupling could be 
inferred by retrofitting experimental data to models like those we present 
here. Since this data would be obtained by purely electrical means via a 
polarized spin current, it could be considerably easier to collect than 
neutron scattering results. Such a method could be a powerful complement to 
current experimental procedures, but is intractable without an understanding 
of the spin wave response to surface STT akin to that which we have outlined
above. In any case, such a scheme would require considerable refinement to
what we have presented here; one would want to keep higher order terms, introduce
another thin-film boundary, and break translational invariance along the surface.
Treating non-single-crystal AFMs would introduce even more complication.
We leave these details to future research, noting here only that continual
improvement of our understanding of AFM spin waves should begin to open new 
routes to experimental investigation on the topic.

This work was supported by the National Science Foundation, Office of
Emerging Frontiers in Research and Innovation EFRI-1433496 (M.W.D), the U.S.
Department of Energy, Office of Basic Energy Sciences, Materials Sciences and
Engineering Division (D.X. and G.M.S.), and by the special funds for the
Major State Basic Research Project of China (grants No. 2014CB921600, No.
2011CB925601) and the National Natural Science Foundation of China (grant No.
91121002) (W.G. and J.X.).

\appendix

\section{Spin-transfer torque on AFM}
\label{sec:Appendix-LLG}

In Sec.~\ref{sec:Macrospin-Model} we presented an extended LLG equation of
motion which included a spin-transfer torque term induced by a 
$\hat{z}$-polarized spin current:
$\tau_\pm = \omega_{s}\mathbf{m}_{\pm}\times\left(\hat{z}\times\mathbf{m}_{\pm}\right)$.
This form of term is plausible on the grounds of right-hand-rule gymnastics, 
but in this appendix we provide a more rigorous derivation of its physical content.

We begin from Eqs.~(6) of Ref. \onlinecite{cheng_spin_2014}, which provide the STT on the $\mathbf{m}$ (magnetization) and $\mathbf{n}$ (staggered) sublattices due to an applied spin voltage\textbf{ $\mathbf{V}_{s}$},
\begin{align}
\tau_{n} & =-\frac{a^{3}}{e\mathcal{V}}G_{r}\mathbf{n}\times\left(\mathbf{m}\times\mathbf{V}_{s}\right)\label{eq:tau-n}\\
\tau_{m} & =-\frac{a^{3}}{e\mathcal{V}}G_{r}\mathbf{n}\times\left(\mathbf{n}\times\mathbf{V}_{s}\right)\label{eq:tau-m}
\end{align}
where $\mathcal{V}$ is the volume of the system, $a$ is the lattice constant, and $G_{r}$ is the real part of the spin mixing conductance for an NM$|$AFM interface; the corresponding imaginary part of $G$ is several orders of magnitudes smaller \cite{cheng_spin_2014} and consequently ignored. By definition, we have $\mathbf{m}_{\pm}=\mathbf{m}\pm\mathbf{n}$ on the two sublattices from Sec. \ref{sec:Macrospin-Model}.  Thus $\tau_{\pm}=\tau_{m}\pm\tau_{n}$, and the use of Eqs.~(\ref{eq:tau-n},\ref{eq:tau-m}) gives
\begin{align}
\tau_{\pm} & =-\frac{a^{3}}{e\mathcal{V}} \mathbf{n}\times
\left(\pm G_{r}\mathbf{m}_{\pm}\times\mathbf{V}_{s}\right).
\label{eq:tau-pm}
\end{align}
Now, as in the main text, we take the spin voltage to be collinear with the easy-axis $\hat{z}$: $\mathbf{V}_s = V_s \hat{z}$. Allowing $\mathbf{n} \approx 2\mathbf{m}_+ \approx -2\mathbf{m}_-$, we have
\begin{equation}
\tau_{\pm} = {a^3V_s\over e\mathcal{V}}G_r
\mathbf{m}_{\pm}\times\left(\hat{z}\times\mathbf{m}_{\pm}\right),
\label{eq:tau-pm-omega_S-form}
\end{equation}
and we define the relevant constant of proportionality as $\omega_{s} = (a^3V_s/e\mathcal{V})G_r$, thus achieving the form exhibited in equation Eq.~\eqref{eq:LLG-macrospin}.

\end{document}